        \documentstyle[12pt]{article}
    \title{{\bf Vertex operator algebras and operads}}
    \author{Yi-Zhi Huang and James Lepowsky}
    \date{{\it This paper is dedicated to the memory of Lawrence Corwin,}\\
	 {\it January 20, 1943 -- March 19, 1992.}}
    
\begin{document}
    \bibliographystyle{alpha}
    \maketitle

    \newtheorem{rema}{Remark}
    \newtheorem{propo}[rema]{Proposition}
    \newtheorem{theo}[rema]{Theorem}
   \newtheorem{defi}[rema]{Definition}
    \newtheorem{lemma}[rema]{Lemma}
    \newtheorem{corol}[rema]{Corollary}
     \newtheorem{exam}[rema]{Example}
  	\newcommand{\ba}{\begin{array}}
	\newcommand{\ea}{\end{array}}
        \newcommand{\be}{\begin{equation}}
        \newcommand{\ee}{\end{equation}}
	\newcommand{\bea}{\begin{eqnarray}}
	\newcommand{\eea}{\end{eqnarray}}
	\newcommand{\nno}{\nonumber}
	\newcommand{\lbar}{\mbox{\huge $\vert$}}
	\newcommand{\p}{\partial}
	\newcommand{\dps}{\displaystyle}

\begin{abstract}
Vertex operator algebras are mathematically rigorous objects
corresponding to chiral algebras in conformal field theory. Operads
are mathematical devices to describe operations, that is, $n$-ary
operations for all $n$ greater than or equal to $0$, not just binary
products. In this paper, a reformulation of the notion of vertex
operator algebra in terms of operads is presented. This reformulation
shows that the rich geometric structure revealed in the study of
conformal field theory and the rich algebraic structure of the theory
of vertex operator algebras share a precise common foundation in basic
operations associated with a certain kind of (two-dimensional)
``complex'' geometric object, in the sense in which classical
algebraic structures (groups, algebras, Lie algebras and the like) are
always implicitly based on (one-dimensional) ``real'' geometric
objects. In effect, the standard analogy between point-particle theory
and string theory is being shown to manifest itself at a more
fundamental mathematical level.
\end{abstract}

\section{Introduction}

The notion of vertex operator algebra arose in the vertex operator
construction of the Monster (see \cite{FLM1}, \cite{B1} and
\cite{FLM2}).  The algebraic theory of vertex operator algebras
provides deep conceptual understanding of many (but not yet all) of
the phenomena of monstrous moonshine (see \cite{CN}) and at the same
time establishes a solid foundation for the algebraic aspects of
conformal field theory.

Though this algebraic theory is mathematically rigorous and is very
powerful in the study of issues ranging from monstrous moonshine, as
in \cite{B2}, \cite{D} and \cite{DM}, to the Batalin-Vilkovisky
algebra structure and the homotopy Lie algebra structure in string
theory (see \cite{LZ}), these results still present certain mysteries.
We propose that many such mysteries would dissipate with a more
conceptual understanding of vertex operator algebras. Early
investigations of the rich geometric structure of conformal field
theory (see especially \cite{BPZ}, \cite{FS},
\cite{Se} and \cite{V}) already suggest that such an
understanding should be related to the geometry of Riemann surfaces.

In \cite{H1} and \cite{H2}, it has been established that the category
of vertex operator algebras, in the sense of \cite{FLM2} or
\cite{FHL}, is isomorphic to a certain category -- the category of
geometric vertex operator algebras -- defined in terms of a certain
moduli space of spheres with punctures and local coordinates. Todd
Trimble and Jim Stasheff commented, in connection with \cite{H1} and
\cite{H3}, respectively, that an operad-like structure appeared to be
implicit in this geometric interpretation of vertex operator algebras.
These comments led to the present reformulation of the notion of
vertex operator algebra using the language of operads; see also our
more detailed paper \cite{HL2}.

The first important example of operad-like structures occurred in
Stasheff's notion of $A_{\infty}$-space, in the course of a
homotopy-theoretic characterization of loop spaces (\cite{St1},
\cite{St2}). Later, May formalized the notion of operad
\cite{M}. But operads can be found ``everywhere'':
Classical algebraic structures (groups, algebras, Lie algebras and so
on) are in fact always implicitly based on operads defined using
one-dimensional geometric objects such as punctured circles and binary
trees, as is pointed out in \cite{H1} and \cite{HL2}. The present
operadic formulation of the notion of vertex operator algebra shows
that vertex operator algebras can be thought of as analogues of
certain classical algebraic structures, in the sense that they too can
be based on certain (partial) operads. The main difference is that for
vertex operator algebras the underlying operads are defined using
certain (two-dimensional) ``complex'' geometric objects instead of
(one-dimensional) ``real'' ones.

Starting from this operadic formulation of the notion of vertex
operator algebra, one can develop the whole theory of vertex operator
algebras and related structures and concepts in parallel with
traditional theories. For instance, the theory of tensor products of
modules for a general class of vertex operator algebras, as developed
beginning in \cite{HL1}, can be expressed using the language of
operads.  And it now appears that this view of the theory of vertex
operator algebras can help us gain a new level of insight into the
phenomena of monstrous moonshine.

In this paper we describe our operadic reformulation. For more
details, see \cite{HL2}. We begin with the definitions of vertex
operator algebra and of operad.  For the basic algebraic theory of
vertex operator algebras, see especially \cite{FLM2} and \cite{FHL}.
Operads and related concepts were introduced in \cite{M}. The details
of the structures and of the proofs of the results in Section 5 are
given in \cite{H1} and \cite{H4}.

We are very grateful to Todd Trimble and to Jim Stasheff for their
illuminating comments on \cite{H1} and \cite{H3}, respectively. We
would also like to thank Israel Gelfand for his interest in this work.
Y.-Z. H. is partially supported by NSF grant DMS-9104519, and J.  L.
by NSF grant DMS-9111945. J. L. also thanks the Institute for Advanced
Study for its hospitality.

\section{The notion of vertex operator algebra}

In this section, we quote the definition and basic ``duality''
properties of vertex operator algebras from \cite{FLM2} or \cite{FHL}.
In this definition, all the variables $x$, $x_{0}, \dots$ are
independent commuting formal variables, and all expressions involving
these variables are to be understood as formal Laurent series. We use
the formal expansion $\delta(x)=\sum_{n\in {\bf Z}}x^{n}$ in the
following way:
\be
x_{0}^{-1}\delta\left(\frac{x_{1}-x_{2}}{x_{0}}\right)=\sum_{n\in {\bf Z}}
\frac{(x_{1}-x_{2})^{n}}{x_{0}^{n+1}}=\sum_{m\in {\bf N},\; n\in {\bf Z}}
(-1)^{m}\left(^{n}_{m}\right)x_{0}^{-n-1}x_{1}^{n-m}x_{2}^{m}.
\ee
(In general, negative powers of binomials are to be expanded in
nonnegative powers of the second summand.)

\begin{defi}
{\rm A {\it vertex operator algebra} (over ${\bf C}$) is a ${\bf
Z}$-graded vector space
\begin{equation}
V=\coprod_{n\in {\bf Z}}V_{(n)}
\end{equation}
such that
\begin{equation}
\mbox{\rm dim }V_{(n)}<\infty\;\;\mbox{\rm for}\; n \in {\bf Z},
\end{equation}
\begin{equation}
V_{(n)}=0\;\;\mbox{\rm for} \;n\; \mbox{\rm sufficiently small},
\end{equation}
equipped with a linear map  $V\otimes V\longrightarrow V[[x, x^{-1}]]$, or
equivalently,
\begin{eqnarray}
&&V\longrightarrow(\mbox{\rm End}\; V)[[x, x^{-1}]]\nonumber \\
&&v\longmapsto Y(v, x)={\displaystyle \sum_{n\in{\bf Z}}}v_{n}x^{-n-1}
\;\;(\mbox{\rm where}\; v_{n}\in
\mbox{\rm End} \;V),
\end{eqnarray}
$Y(v, x)$ denoting the {\it vertex operator associated with} $v$, and
equipped also with two distinguished homogeneous vectors ${\bf 1}\in
V_{(0)}$ (the {\it vacuum}) and $\omega \in V_{(2)}$. The following
conditions are assumed for $u, v \in V$:
\begin{equation}
u_{n}v=0\;\;\mbox{\rm for}\;n\; \mbox{\rm sufficiently large};
\end{equation}
\begin{equation}
Y({\bf 1}, x)=1\;\; (1\;\mbox{\rm on the right being the identity
operator});
\end{equation}
the {\it creation property} holds:
\begin{equation}
Y(v, x){\bf 1} \in V[[x]]\;\;\mbox{\rm and}\;\;\lim_{x\rightarrow
0}Y(v, x){\bf 1}=v
\end{equation}
(that is, $Y(v, x){\bf 1}$ involves only nonnegative integral powers
of $x$ and the constant term is $v$); the {\it Jacobi identity} (the
main axiom) holds:
\begin{eqnarray}
&x_{0}^{-1}\delta
\left({\displaystyle\frac{x_{1}-x_{2}}{x_{0}}}\right)Y(u, x_{1})Y(v,
x_{2})-x_{0}^{-1} \delta
\left({\displaystyle\frac{x_{2}-x_{1}}{-x_{0}}}\right)Y(v, x_{2})Y(u,
x_{1})&\nonumber \\ &=x_{2}^{-1} \delta
\left({\displaystyle\frac{x_{1}-x_{0}}{x_{2}}}\right)Y(Y(u, x_{0})v,
x_{2})&
\end{eqnarray}
(note that when each expression in (9) is applied to any element of
$V$, the coefficient of each monomial in the formal variables is a
finite sum; on the right-hand side, the notation $Y(\cdot, x_{2})$ is
understood to be extended in the obvious way to $V[[x_{0},
x^{-1}_{0}]]$); the Virasoro algebra relations hold:
\begin{equation}
[L(m), L(n)]=(m-n)L(m+n)+{\displaystyle\frac{1}{12}}
(m^{3}-m)\delta_{n+m,0}c
\end{equation}
for $m, n \in {\bf Z}$, where
\begin{equation}
L(n)=\omega _{n+1}\;\; \mbox{\rm for} \;n\in{\bf Z}, \;\;{\rm
i.e.},\;\;Y(\omega, x)=\sum_{n\in{\bf Z}}L(n)x^{-n-2}
\end{equation}
and
\begin{eqnarray}
&c\in {\bf C};&\\ &L(0)v=nv\;\;\mbox{\rm for}\;n \in {\bf
Z}\;\mbox{\rm and}\;v\in V_{(n)};&\\ &{\displaystyle \frac{d}{dx}}Y(v,
x)=Y(L(-1), x).&
\end{eqnarray}}
\end{defi}

The vertex operator algebra just defined is denoted by $(V, Y, {\bf 1},
\omega)$. The complex number $c$ is called the {\it central charge}
or {\it rank} of $V$ (or of $(V, Y, {\bf 1}, \omega)$).

Vertex operator algebras have important ``rationality,''
``commutativity'' and ``associativity'' properties, collectively
called ``duality'' properties.  These properties in fact can be used
as axioms replacing the Jacobi identity in the definition of vertex
operator algebra, as we now explain.

In the propositions below, ${\bf C}[x_{1}, x_{2}]_{S}$ is the ring of
rational functions obtained by inverting (localizing with respect to)
the products of (zero or more) elements of the set $S$ of nonzero
homogeneous linear polynomials in $x_{1}$ and $x_{2}$. Also,
$\iota_{12}$ is the operation of expanding an element of ${\bf
C}[x_{1}, x_{2}]_{S}$, that is, a polynomial in $x_{1}$ and $x_{2}$
divided by a product of linear polynomials in $x_{1}$ and $x_{2}$, as
a formal series containing at most finitely many negative powers of
$x_{2}$ (using binomial expansions for negative powers of linear
polynomials involving both $x_{1}$ and $x_{2}$); similarly for
$\iota_{21}$ and so on. (The distinction between rational functions
and formal Laurent series is crucial.)
\begin{propo}
{\bf (a) (rationality of products)} For $v$, $v_{1}$, $v_{2}\in V$ and
$v'\in V'$ (the graded dual space of $V$), the formal series
$\left\langle v', Y(v_{1}, x_{1})Y(v_{2}, x_{2})v\right\rangle,$ which
involves only finitely many negative powers of $x_{2}$ and only
finitely many positive powers of $x_{1}$, lies in the image of the map
$\iota_{12}$:
\begin{equation}
\left\langle v', Y(v_{1}, x_{1})Y(v_{2}, x_{2})v\right\rangle
=\iota_{12}f(x_{1}, x_{2}),
\end{equation}
where the (uniquely determined) element $f\in {\bf C}[x_{1},
x_{2}]_{S}$ is of the form
\begin{equation}
f(x_{1}, x_{2})={\displaystyle \frac{g(x_{1},
x_{2})}{x_{1}^{r}x_{2}^{s}(x_{1}-x_{2})^{t}}}
\end{equation}
for some $g\in {\bf C}[x_{1}, x_{2}]$ and $r, s, t\in {\bf Z}$.

{\bf (b) (commutativity)} We also have
\begin{equation}
\left\langle v', Y(v_{2}, x_{2})Y(v_{1}, x_{1})v\right\rangle
=\iota_{21}f(x_{1}, x_{2}).
\end{equation}
\end{propo}
\begin{propo}
{\bf (a) (rationality of iterates)} For $v$, $v_{1}$, $v_{2}\in V$ and
$v'\in V'$, the formal series
$\left\langle v', Y(Y(v_{1}, x_{0})v_{2}, x_{2})v\right\rangle,$
which involves only finitely many negative powers of $x_{0}$ and only
finitely many positive powers of $x_{2}$, lies in the image of the map
$\iota_{20}$:
\begin{equation}
\left\langle v', Y(Y(v_{1}, x_{0})v_{2},
x_{2})v\right\rangle=\iota_{20}h(x_{0}, x_{2}),
\end{equation}
where the (uniquely determined) element $h\in {\bf C}[x_{0},
x_{2}]_{S}$ is of the form
\begin{equation}
h(x_{0}, x_{2})={\displaystyle \frac{k(x_{0},
x_{2})}{x_{0}^{r}x_{2}^{s}(x_{0}+x_{2})^{t}}}
\end{equation}
for some $k\in {\bf C}[x_{0}, x_{2}]$ and $r, s, t\in {\bf Z}$.

{\bf (b)} The formal series
$\left\langle v', Y(v_{1}, x_{0}+x_{2})Y(v_{2}, x_{2})v\right\rangle,$
which involves only finitely many negative powers of $x_{2}$ and only
finitely many positive powers of $x_{0}$, lies in the image of
$\iota_{02}$, and in fact
\begin{equation}
\left\langle v', Y(v_{1}, x_{0}+x_{2})Y(v_{2},
x_{2})v\right\rangle=\iota_{02}h(x_{0}, x_{2}).
\end{equation}
\end{propo}
\begin{propo}[associativity]
We have the following equality of rational functions:
\begin{equation}
\iota_{12}^{-1}\left\langle v', Y(v_{1}, x_{1})Y(v_{2}, x_{2})v\right\rangle
=(\iota_{20}^{-1}\left\langle v', Y(Y(v_{1}, x_{0})v_{2},
x_{2})v\right\rangle)\lbar_{x_{0}=x_{1}-x_{2}}.
\end{equation}
\end{propo}
\begin{propo}
The Jacobi identity follows from the rationality of products and
iterates, and commutativity and associativity. In particular, in the
definition of vertex operator algebra, the Jacobi identity may be
replaced by these properties.
\end{propo}

These duality properties, when suitably interpreted, form part of the
geometric interpretation of vertex operator algebras discussed in
Section 5 below. This part of the geometric interpretation was
initially pointed out by Igor Frenkel.

\section{The notions of operads and their modules}

Now we give a variant of May's definition of operad (cf. \cite{M}):

\begin{defi}
{\rm An {\it operad} ${\cal C}$ consists of a family of sets ${\cal
C}(j)$, $j\in {\bf N}$, together with {\it (abstract) substitution
maps} $\gamma$, one for each $k\in {\bf N}$, $j_{1}, \dots, j_{k}\in
{\bf N}$,
\begin{eqnarray}
\gamma:\; {\cal C}(k)\times {\cal C}(j_{1})\times \cdots\times
{\cal C}(j_{k})&\longrightarrow &{\cal C}(j_{1}+\cdots
+j_{k})\nonumber\\ (c; d_{1}, \dots, d_{k})&\longmapsto& \gamma(c;
d_{1}, \dots, d_{k}),
\end{eqnarray}
an {\it identity element} $I\in {\cal C}(1)$ and a (left) action of
the symmetric group $S_{j}$ on ${\cal C}(j)$, $j\in {\bf N}$ (where
$S_0$ is understood to be the trivial group), satisfying the following
axioms:

(i) {\it Operad-associativity}: For any $k\in {\bf N}$, $j_{s}\in {\bf
N}$ ($s=1,
\dots, k$), $i_{t}\in {\bf N}$ ($t=1, \dots, j_{1}+\cdots +j_{k}$),
$c\in {\cal C}(k)$, $d_{s}\in {\cal C}(j_{s})$ ($s=1,,\dots, k$) and
$e_{t}\in {\cal C}(i_{t})$ ($t=1,\dots,j_{1}+\cdots +j_{k}$),
\be
\gamma(\gamma(c;d_{1},\dots,d_{k});e_{1},\dots,e_{j_{1}+\cdots +j_{k}})\\
=\gamma(c;f_{1}, \dots, f_{k}),
\ee
where
\be
f_{s}=\gamma(d_{s};e_{j_{1}+\cdots +j_{s-1}+1},\dots,e_{j_{1}+\cdots +j_{s}}).
\ee

(ii) For any $j,k\in {\bf N}$, $d\in {\cal C}(j)$ and $c\in {\cal C}(k)$,
\be
\gamma(I;d)=d,
\ee
\be
\gamma(c;I,\dots, I)=c.
\ee
(In particular, for $k=0$, $\gamma: {\cal C}(0)\longrightarrow {\cal C}(0)$
is the identity map.)

(iii) For any $k\in {\bf N}$, $j_{s}\in {\bf N}$ ($s=1, \dots,
k$), $c\in {\cal C}(k)$, $d_{s}\in{\cal C}(j_{s})$ ($s=1, \dots, k$),
$\sigma\in S_{k}$ and $\tau_{s}\in S_{j_{s}}$ ($s=1, \dots, k$),
\be
\gamma(\sigma(c);d_{1},\dots,d_{k})=\sigma(j_{1}, \dots,
j_{k})(\gamma(c;d_{\sigma(1)},\dots,d_{\sigma(k)})),
\ee
\be
\gamma(c;\tau_{1}(d_{1}),\dots,\tau_{k}(d_{k}))=(\tau_{1}\oplus
\cdots\oplus\tau_{k})(\gamma(c;d_{1},\dots,d_{k})),
\ee
where $\sigma(j_{1}, \dots, j_{k})$ denotes the permutation of
$j=\sum_{s=1}^{k}j_{s}$ letters which permutes the $k$ blocks of
letters determined by the given partition of $j$ as $\sigma$ permutes
$k$ letters, and $\tau_{1}\oplus\cdots\oplus\tau_{k}$ denotes the
image of $(\tau_{1},\dots, \tau_{k})$ under the obvious inclusion of
$S_{j_{1}}\times \cdots\times S_{j_{k}}$ in $S_{j}$; that is,
\be
\sigma(j_{1}, \dots, j_{k})(j_{\sigma(1)}+ \cdots +j_{\sigma(i-1)}+
l)=j_{1}+\cdots +j_{\sigma(i)-1}+l, \;\;l=1, \dots, j_{\sigma(i)},
\;\; i=1, \dots, k
\ee
and
\be
(\tau_{1}\oplus\cdots\oplus\tau_{k})(j_{1}+\cdots +j_{i-1}+l)=j_{1}+\cdots
+j_{i-1}+\tau_{i}(l),\;\;l=1,\dots, j_{i},\;\;i=1,\dots,k.
\ee
(May \cite{M} uses right actions, so that the description of the
permutation $\sigma(j_{1}, \dots, j_{k})$ might have a different
interpretation.)}
\end{defi}

This definition is the same as that in \cite{M} with the
following exceptions:

(i) ${\cal C}(0)$ need not consist of exactly one element.

(ii) The sets in the definition need not be (certain kinds of)
topological spaces (and correspondingly the maps need not be
continuous).

\begin{defi}
{\rm Suppose that in the definition above, the substitution maps
$\gamma$ are only partially defined, that is, each map $\gamma$ in
formula (22) takes a subset of ${\cal C}(k)\times {\cal
C}(j_{1})\times \cdots\times{\cal C}(j_{k})$ to ${\cal C}(j_{1}+\cdots
+j_{k})$; all the other data remain the same; each of the formulas
(23) and (25) -- (28) holds whenever {\it both} sides exist; and the
left-hand sides of (25) and (26) always
exist. Then we call such a family of sets ${\cal C}(j)$ together with
the partial maps $\gamma$, the identity $I$ and the actions of $S_{j}$
on ${\cal C}(j)$ a {\it partial operad}. In addition, we define a {\it partial
pseudo-operad} to be a family of sets ${\cal C}(j)$, $j\in {\bf N}$,
together with partially defined substitution maps $\gamma$, an
identity $I$ and actions of $S_{j}$ on ${\cal C}(j)$, $j\in {\bf N}$,
satisfying all the axioms for partial operads except the
operad-associativity.  (Later, we shall typically denote partial
operads by the symbol ${\cal P}$ rather than ${\cal C}$.)}
\end{defi}

If in Definitions 6 and 7 the sets ${\cal C}(j)$, $j\in
{\bf N}$, are assumed to be objects in given categories (e.g., have
certain kinds of topological, smooth or analytic structure) and the
maps $\gamma$ and the actions of $S_{j}$, $j\in {\bf N}$, are
morphisms in these categories (e.g., are continuous or smooth or
analytic), we have the notions of operads in these categories, and we
use the names of these categories plus the word ``operads" to
designate them (e.g., {\it topological operads, smooth operads} or
{\it analytic operads}). In the case of partial operads, we also
require that the domains of the substitution maps are in the category
we are considering.

A {\it morphism} $\psi: {\cal C} \longrightarrow {\cal C}'$ of operads
${\cal C}$ and ${\cal C}'$ is a sequence of $S_{j}$-equivariant maps
$\psi_{j}: {\cal C}(j)\longrightarrow {\cal C}'(j)$ such that
$\psi_{1}(I)=I'$ and the following diagram commutes:
\be
\begin{array}{ccc}
{\cal C}(k)\times{\cal C}(j_{1})\times\cdots\times{\cal
C}(j_{k})&\stackrel{\gamma}{\longrightarrow}&{\cal C}(j_{1}+\cdots
+j_{k})\\
\downarrow&&\downarrow\\
{\cal C}'(k)\times{\cal C}'(j_{1})\times\cdots\times{\cal
C}'(j_{k})&\stackrel{\gamma'}{\longrightarrow}&{\cal C}'(j_{1}+\cdots
+j_{k}).
\end{array}
\ee

For partial operads we also require that the domains of the
substitution maps for ${\cal C}$ are mapped into the domains of the
substitution maps for ${\cal C}'$; the diagram (31) is interpreted in
the obvious way.  Morphisms for partial pseudo-operads are
defined in the same way as morphisms for partial operads.

Next we discuss the sense in which operads describe ``operations.'' In
the rest of this section, all the operads will be ordinary (i.e., in
the category of sets and not partial).

Let $X$ be a set and $Y$ a subset of $X$. We define the {\it
endomorphism operad} ${\cal E}_{X,Y}$ as follows: Let
$*$ be a one-element set (a terminal object in the category of sets)
and take $X^0=Y^0=*$. Let ${\cal E}_{X,Y}(j)$, $j\in {\bf N}$, be the
set of maps from $X^{j}$ to $X$ which map $Y^{j}$ to $Y$; then ${\cal
E}_{X, Y}(0)=Y$.  The substitution maps are defined by
\be
\gamma(f;g_{1}, \dots, g_{k})=f\circ (g_{1}\times \cdots\times g_{k})
\ee
for $f\in {\cal E}_{X,Y}(k)$, $k\in {\bf N}$ and $g_{s}\in {\cal
E}_{X,Y}(j_{s})$, $s=1, \dots, k$. The identity $I_{X,Y}$ is the
identity map of $X$. For $f\in {\cal E}_{X,Y}(j)$, $\sigma\in S_{j}$,
${\bf x}=(x_{1}, \dots, x_{j})\in X^{j}$,
\be
(\sigma(f))({\bf x})=f(\sigma^{-1}({\bf x}))
\ee
where
\be
\sigma({\bf x})=\sigma(x_{1}, \dots, x_{j})=(x_{\sigma^{-1}(1)}, \dots,
x_{\sigma^{-1}(j)}).
\ee
It is easy to see that ${\cal E}_{X,Y}$ is an operad. The
corresponding definition in \cite{M} amounts to the case in which $Y$
has one element. Note the special cases $Y= \emptyset$ and $Y=X$.
Observe that operad-associativity (the associativity of substitution)
is unrelated to any associativity properties that the $j$-ary
operations might or might not have.

For vector spaces we have a more appropriate notion. Let $V$ be a
vector space and $W$ a subspace of $V$.  We define the corresponding
{\it (multilinear) endomorphism operad} ${\cal M}_{V,W}$ as follows:
Let ${\cal M}_{V,W}(j)$, $j\in {\bf N}$, be the set of multilinear
maps from $V^{j}$ to $V$ which map $W^{j}$ to $W$; it is understood
that $V^0=W^{0}$ is the one-element set as above, and that a
``multilinear map" (a ``zero-linear map") from this set to $V$ (or
$W$) is a map of sets, i.e., an element of the target set. In
particular, ${\cal M}_{V, W}(0)=W$.  The substitution maps, the
identity and the actions of the symmetric groups are defined just as
in the definition of endomorphism operads for sets.  Then ${\cal
M}_{V,W}$ is an operad. Note the special cases $W=0$ and $W=V$.

\begin{defi}
{\rm Let ${\cal C}$ be an operad. A {\it ${\cal C}$-space} consists of
a set $X$, a subset $Y\subset X$ and a morphism $\psi$ of operads from
${\cal C}$ to ${\cal E}_{X,Y}$ such that $\psi_{0}({\cal C}(0))={\cal
E}_{X,Y}(0)\;(=Y)$. It is denoted $(X,Y, \psi)$.  An element of
$\psi_{0}({\cal C}(0))=Y$ is called a {\it quasi-identity element of
$X$ (for ${\cal C}$ and $\psi$)}. Note that each element of ${\cal
C}(j)$, $j\in {\bf N}$, defines a $j$-ary operation on $X$. A {\it
morphism} from a ${\cal C}$-space $(X, Y, \psi)$ to a ${\cal C}$-space
$(X', Y', \psi')$ is a map $\eta: X\longrightarrow X'$ such that
$\eta(Y)\subset Y'$ and $\eta \circ \psi_j(c)=\psi'_j(c) \circ \eta^j$
for $j \in {\bf N}$ and $c \in {\cal C}(j)$; it follows that
$\eta(Y)=Y'$. An {\it isomorphism} of ${\cal C}$-spaces is defined in
the obvious way.  The definition of ``${\cal C}$-space" in \cite{M}
amounts to the case in which ${\cal C}(0)$ and $Y$ consist of one
element.}
\end{defi}

\begin{defi}
{\rm A {\it ${\cal C}$-module} $(V, W, \nu)$ consists of a vector
space $V$, a subspace $W$ and a morphism $\nu$ from ${\cal C}$ to
${\cal M}_{V, W}$ such that the subspace of $V$ spanned by
$\nu_{0}({\cal C}(0))$ is ${\cal M}_{V,W}(0)\;(=W)$. We call
$\nu_{0}({\cal C}(0))$ the set of {\it quasi-identity elements}. Each
element of ${\cal C}(j)$, $j\in {\bf N}$, defines a multilinear
$j$-ary operation on $V$.  Morphisms and  isomorphisms of
${\cal C}$-modules are defined in the obvious ways. }
\end{defi}

The notion of {\it suboperad} of an operad is defined in the obvious
way.  An intersection of suboperads is again a suboperad. Let ${\cal
C}$ be an operad and $U$ a subset of the disjoint union of the sets
${\cal C}(j)$, $j\in {\bf N}$. The {\it suboperad of ${\cal C}$
generated by $U$} is the smallest suboperad of ${\cal C}$ such that
the disjoint union of the family of sets in the suboperad contains
$U$. The operad ${\cal C}$ is said to be {\it generated by $U$} if the
suboperad generated by $U$ is ${\cal C}$ itself.

\begin{defi}
{\rm Let ${\cal C}$ be an operad. We call an element $a$ of ${\cal
C}(2)$ {\it associative} if
\be
\gamma(a; a, I)
=\gamma(a; I, a).
\ee
We say that an operad ${\cal C}$ is  {\it associative} if
${\cal C}$ is generated by ${\cal C}(0)$ and an associative element
$a\in {\cal C}(2)$.}
\end{defi}

\begin{defi}
{\rm Let ${\cal C}$ be an associative operad with associative element
$a\in {\cal C}(2)$. We call a ${\cal C}$-space a {\it ${\cal
C}$-monoid} and a ${\cal C}$-module a {\it ${\cal C}$-associative
algebra}, with {\it ${\cal C}$-associative binary product} given by
the image of $a$ and with {\it quasi-identity elements} given by the
image of ${\cal C}(0)$.}
\end{defi}

An important example of an associative operad can be obtained from
circles with punctures and local coordinates. We consider the moduli
spaces $C(j)$, $j\in {\bf N}$, of circles (i.e., compact connected
smooth one-dimensional manifolds) with $j+1$ ordered points (called
{\it punctures}), the zeroth negatively oriented, the others
positively oriented, and with local coordinates vanishing at these
punctures.  Given two such circles with punctures and local
coordinates and a positively oriented puncture on the first circle, we
can cut suitable intervals around the given puncture and the
negatively oriented puncture on the second circle, and then identify
their boundaries using the two local coordinate maps and the map
$t\longmapsto 1/t$ to obtain another circle with punctures and local
coordinates. The ordering of the positively oriented punctures of the
sewn circle is obtained by ``inserting'' the ordering for the second
circle into that for the first. This procedure in fact induces an
operation on the moduli space of circles with punctures and local
coordinates. Using this operation on the moduli space we can define
substitution maps. We also have a natural identity element and actions
of $S_{j}$ on the $C(j)$, $j\in {\bf N}$. The family $\{C(j)\;|\;j\in {\bf
N}\}$ with these structures forms an associative operad, which we
denote by $C$. For more details, see
\cite{HL2}.

For this operad $C$, a $C$-monoid is a monoid in the usual sense and a
$C$-associative algebra is an associative algebra in the usual sense.

\section{Rescalable partial operads and their modules}

Let ${\cal P}$ be a partial operad. A subset $G$ of ${\cal P}(1)$ is
called a {\it rescaling group} for ${\cal P}$ if $G$ contains $I$; the
substitution maps $\gamma$ from a subset of ${\cal P}(1)\times {\cal
P}(k)$ to ${\cal P}(k)$ and from a subset of ${\cal P}(k)\times ({\cal
P}(1))^{k}$ to ${\cal P}(k)$ are defined on $G\times {\cal P}(k)$ and
on ${\cal P}(k)\times G^{k}$, respectively, for each $k\in {\bf N}$;
both sides of (23) exist if $c\in G$ or $d_{1}, \dots, d_{k}\in G$ or
$e_{1}, \dots, e_{j_{1}+\cdots +j_{k}}\in G$ and if either side of (23)
exists; $\gamma$ maps $G\times G$ into $G$; and inverses of the elements of
$G$ exist with respect to $\gamma$ and $I$; then $G$ is in fact a
group. (Note that $G=\{ I\}$ is always an example of a rescaling group
for ${\cal P}$.) Given a rescaling group $G$ for ${\cal P}$, we define
a corresponding equivalence relation on ${\cal P}$: Two elements
$c_{1}$ and $c_{2}$ of ${\cal P}(j)$, $j\in {\bf N}$, are said to be
{\it $G$-equivalent} if there exists $d\in G$ such that
\be
c_{2}=\gamma(d; c_{1});
\ee
our assumptions insure that this is an equivalence
relation.

\begin{defi}
{\rm A {\it ($G$-)rescalable partial operad} is a partial operad
${\cal P}$ together with a rescaling group $G$ for ${\cal P}$
satisfying the following condition: For any $c\in {\cal P}(k)$, $k\in
{\bf N}$, $d_{1}\in {\cal P}(j_{1}), \dots, d_{k}\in {\cal P}(j_{k})$,
$j_{i}\in {\bf N}$, there exist $d'_{1}\in {\cal P}(j_{1}),
\dots,d'_{k}\in {\cal P}(j_{k})$ which are $G$-equivalent to $d_{1},
\dots,d_{k}$, respectively, such that $\gamma(c; d'_{1}, \dots,
d'_{k})$ exists.}
\end{defi}

Suppose that we have a set-theoretic category with a reasonable notion
of ``induced substructure," such as a topological, smooth or analytic
category. A {\it partial operad with rescaling group} (or {\it
rescalable partial operad}) {\it in this category} (e.g., a {\it
topological}, {\it smooth} or {\it analytic rescalable partial
operad}) is a partial operad ${\cal P}$ with rescaling group (or a
rescalable partial operad) such that its underlying partial operad is
in the category we are considering and the rescaling group is a group
in this category, with the structure induced from that on ${\cal
P}(1)$.

Morphisms and isomorphisms of partial
operads with rescaling groups are defined in the obvious ways.

The definitions of ${\cal P}$-space and ${\cal P}$-module in Section 3
also make sense when ${\cal P}$ is a partial operad. But for a partial
operad ${\cal P}$ with a rescaling group $G$, it is more relevant to
look for some kind of ``(multilinear) endomorphism partial operad" of a
$G$-module, and then to define a ``${\cal P}$-module" to be a
morphism from ${\cal P}$ to such a (multilinear) endomorphism partial
operad.  However, we must be content with only the following
``(multilinear) endomorphism partial pseudo-operads" (recall Definition 7):

Let $G$ be a group, $V$ a completely reducible $G$-module and $W$ a
$G$-submodule of $V$.  Then $V=\coprod_{M\in A}V_{(M)}$, where $A$ is
the set of equivalence classes of irreducible $G$-modules and
$V_{(M)}$ is the sum of the $G$-submodules of $V$ in the class $M$,
and similarly for $W$.  Assume that dim $V_{(M)}<\infty$ for every
$M\in A$. We define a {\it (multilinear) endomorphism partial
pseudo-operad} ${\cal H}^{G}_{V, W}$ as follows: For any $j\in {\bf
N}$ the set ${\cal H}^{G}_{V,W}(j)$ is the set of all multilinear maps from
$V^{ j}$ to $\overline{V}=\prod_{M\in A}V_{(M)}=V'^{*}$ such that
$W^{j}$ is mapped to $\overline{W}=\prod_{M\in A}W_{(M)}=W'^{*}$,
where $'$ denotes the graded dual of an $A$-graded vector space and
$^{*}$ denotes the dual space of a vector space. As in the definition
of multilinear endomorphism operad above, it is understood that
$V^0=W^0$ is the one-element set, so that the set ${\cal H}^{G}_{V,
W}(0)$ is equal to $\overline{W}$. The identity $I^{G}_{V,W}$ is the
embedding map from $V$ to $\overline{V}$. The symmetric group $S_{j}$
acts on ${\cal H}_{V,W}^{G}(j)$ in the obvious way.  To define the
substitution maps, we first define a contraction operation on ${\cal
H}^{G}_{V,W}$: Given $f\in {\cal H}^{G}_{V,W}(k)$ and $g\in {\cal
H}^{G}_{V,W}(j)$ ($k,j\in {\bf N}$) and a positive integer $s\le k$,
we say that the {\it contraction of $f$ at the $s$-th argument and $g$
at the zeroth argument exists} if for any $v_{1}, \dots, v_{k+j-1}\in
V$ and $v'\in V'$, the series
\be
\sum_{M\in A}\langle v', f(v_{1}, \dots, v_{s-1}, P_{M}(g(v_{s},
\dots, v_{s+j-1})), v_{s+j}, \dots, v_{k+j-1})\rangle
\ee
converges absolutely, where $P_{M}: \overline{V}\longrightarrow
V_{(M)}$ is the projection operator. In this case the (well-defined)
limits for all $v_{1},\dots,v_{k+j-1}\in V$, $v'\in V'$ define an
element $f_{^{\;s\!\!}}*_{^{0}}g$ of ${\cal H}^{G}_{V,W}(k+j-1)$, the
{\it contraction}. More generally, given any subset of $\{ 1, \dots,
k\}$ and any element of $\cup_{j\in {\bf N}}{\cal H}_{V,W}^{G}(j)$ for
each element of the subset, we have the analogous {\it contraction},
defined using the appropriate multisums, when they are absolutely
convergent. The substitution map
\begin{eqnarray}
\gamma^{G}_{V,W}: {\cal H}^{G}_{V,W}(k)\times {\cal H}^{G}_{V,W}(j_{1})
\times \cdots \times {\cal H}^{G}_{V,W}(j_{k})&\longrightarrow
&{\cal H}^{G}_{V,W}(j_{1}+\cdots +j_{k})\nonumber\\
(f; g_{1},\dots, g_{k})&\longmapsto &\gamma^{G}_{V,W}(f; g_{1},\dots, g_{k})
\end{eqnarray}
is defined by this procedure, using the whole set $\{ 1, \dots, k\}$.
Of course, the cases of proper subsets of $\{ 1, \dots, k\}$ are
recovered by letting some of the $g_{l}$ be $I_{V,W}^{G}$.  The family
${\cal H}_{V,W}^{G}$ of sets ${\cal H}_{V,W}^{G}(j)$, $j\in {\bf N}$,
equipped with the substitution maps $\gamma_{V,W}^{G}$, the identity
$I_{V,W}^{G}$ and the actions of $S_{j}$ on ${\cal H}_{V,W}^{G}(j)$,
$j\in {\bf N}$, satisfies all the axioms for a partial operad except
the operad-associativity and therefore is a partial pseudo-operad. The
operad-associativity fails because in general we cannot expect to have
the absolute convergence of the multisums corresponding to a sequence
of substitutions.

Using (multilinear) endomorphism partial pseudo-operads, we define the
following notions of ${\cal P}$-pseudo-module and of ${\cal
P}$-module:

\begin{defi}
{\rm Let ${\cal P}$ be a partial operad with rescaling group $G$. A
{\it ${\cal P}$-pseudo-module} $(V, W, \nu)$ is a completely reducible
$G$-module $V=\coprod_{M\in A}V_{(M)}$ with dim $V_{(M)}< \infty$,
together with a submodule $W$ of $V$ and a morphism $\nu$ from ${\cal
P}$ (viewed as a partial pseudo-operad) to the partial pseudo-operad
${\cal H}^{G}_{V, W}$, such that the submodule of $V$ generated by the
homogeneous components of the elements of $\nu_{0}({\cal P}(0))$ is
$W$ and the map from $G$ to ${\cal H}^{G}_{V, W}(1)$ induced from
$\nu_{1}$ is the given representation of $G$ on $V$.  An element of
$\nu_{0}({\cal P}(0))$ is called a {\it quasi-identity element of
$\overline{V}$ for ${\cal P}$}.  A {\it morphism} from a ${\cal
P}$-pseudo-module $(V, W, \nu)$ to a ${\cal P}$-pseudo-module
$(\tilde{V}, \tilde{W}, \tilde{\nu})$ is a $G$-module morphism $\eta:
V\longrightarrow \tilde{V}$ such that $\eta(W)\subset \tilde{W}$ and
$\overline{\eta}\circ \nu_{j}(c)=\tilde{\nu}_{j}(c)\circ \eta^{j}$ for
$j\in {\bf N}$ and $c\in {\cal P}(j)$, where $\eta$ is extended
naturally to $\overline{\eta}: \overline{V}\longrightarrow
\overline{\tilde{V}}$; it follows that $\eta(W)=\tilde{W}$.
Isomorphisms of ${\cal P}$-pseudo-modules are defined in the obvious
way. For a ${\cal P}$-pseudo-module $(V,W, \nu)$, the image of ${\cal
P}$ under $\nu$ (where it is understood that the substitution maps are
the substitution maps for ${\cal H}_{V,W}^{G}$ restricted to the
images of the domains of the substitution maps for ${\cal P}$) is a
partial pseudo-operad.  We define a {\it ${\cal P}$-module} to be a
${\cal P}$-pseudo-module $(V,W, \nu)$ such that the image of ${\cal
P}$ under $\nu$ is a partial operad, that is, such that
operad-associativity holds for the image. Morphisms and isomorphisms
of ${\cal P}$-modules are defined to be morphisms and isomorphisms of
the underlying ${\cal P}$-pseudo-modules, respectively.}
\end{defi}

Though this definition of ${\cal P}$-module is conceptually
natural, it is in practice typically very difficult to determine
whether a ${\cal P}$-pseudo-module is a ${\cal P}$-module. The issue
is to insure that operad-associativity holds for certain families of
multilinear maps.

The notion of {\it partial
suboperad} of a partial operad is defined in the obvious way; we
require that substitutions in a partial suboperad exist if and only if
the corresponding substitutions  in the original partial operad exist.
An intersection of partial suboperads of a partial operad is a partial
suboperad. We also have the notion of {\it partial suboperad generated
by a subset}. If a partial operad is the partial suboperad
generated by a given subset, we say that this partial operad is {\it
generated by the subset}.

\begin{defi}
{\rm Let ${\cal P}$ be a partial operad with rescaling group $G$. We
call an element $a\in {\cal P}(2)$ {\it associative} if there exists
$a'\in {\cal P}(2)$ which is $G$-equivalent to $a$ (that is, there
exists $b_{0}\in G$ such that $a'=\gamma(b_{0}; a)$) and there exist
unique $b_{i}\in G$, $i=1, \dots, 5$, which depend on $a'$, such that
$\gamma(a; a', I)$ exists and
\be
\gamma(a; a', I)
=\gamma(d_{1}; I, d_{2}),
\ee
where
\be
d_{1}=\gamma(b_{1}; \gamma(a; b_{2},
b_{3})),\;\;\;\;d_{2}=\gamma(a; b_{4}, b_{5}).
\ee
We call a partial operad ${\cal P}$ with rescaling group $G$ {\it
associative} if ${\cal P}$ is generated by ${\cal P}(0)$, $G$ and an
associative element $a\in {\cal P}(2)$.}
\end{defi}

\begin{defi}
{\rm Let ${\cal P}$ be an associative partial operad with rescaling
group $G$ and associative element $a\in {\cal P}(2)$. We call a ${\cal
P}$-pseudo-module a {\it ${\cal P}$-associative pseudo-algebra} and a
${\cal P}$-module a {\it ${\cal P}$-associative algebra}, with {\it
${\cal P}$-associative binary product} given by the image of $a$ and
with {\it quasi-identity elements} given by the image of ${\cal
P}(0)$. }
\end{defi}

\section{Vertex operator algebras as modules for certain partial operads
associated with spheres with tubes}

Using the language developed in the previous section, we can
now explain how vertex operator algebras (recall  Section 2) amount to
modules for the partial operads which are the complex powers of the
determinant line bundle over the moduli space of spheres with tubes.

A {\it sphere with $n$ tubes} ($n>0$) is a sphere (a genus-zero
compact connected one-dimensional complex manifold) with $n$ distinct,
ordered points (called {\it punctures}) with the zeroth puncture
negatively oriented and the other punctures positively oriented, and
with local analytic coordinates vanishing at these punctures. Given
two spheres with tubes and given one positively oriented puncture on
the first sphere, we can sew these two spheres at the given puncture
on the first sphere and the negatively oriented puncture on the second
sphere by first cutting disks with reciprocal radii (using the local
coordinates), and containing no other punctures, around the two given
punctures, and then identifying the boundaries of the remaining parts
of the two spheres with tubes using the two local coordinate maps and
the map $z\longmapsto 1/z$. We call this procedure the {\it sewing
operation}.  Note that the conditions ``with reciprocal radii'' and
``containing no other punctures'' are not always satisfied, therefore
the sewing operation is only a partial operation. But if we rescale
the local coodinate map at the negatively oriented puncture on the
second sphere with tubes by multiplying it by a suitable nonzero
complex number, then these two conditions can always be satisfied, and
thus after rescaling, any two spheres with tubes can always be sewn
together at the given punctures. The ordering of the positively
oriented punctures on the sewn sphere is obtained by ``inserting'' the
ordering for the second sphere into that for the first.

Two spheres with tubes are said to be {\it conformally equivalent} if
there exists an analytic diffeomorphism of the underlying
one-dimensional complex manifolds preserving all the indicated
structures except perhaps the local coordinate neighborhoods. The
space of conformal equivalence classes of spheres with $n$ tubes is
called the {\it moduli space of spheres with $n$ tubes} and is denoted
$K(n-1)$ ($n-1$ being the number of positively oriented punctures).
Given two conformal equivalence classes in $K(m)$ and $K(n)$,
respectively, and a positive integer $i\le m$, if there are two
spheres with tubes in the two given conformal equivalence classes
which can be sewn at the the $i$-th positively oriented puncture on
the first sphere and the negatively oriented puncture on the second
sphere, the conformal equivalence class of the sewn sphere with tubes,
in $K(m+n-1)$, depends only on the two given classes. This procedure
for obtaining an element of $K(m+n-1)$ from two elements of $K(m)$ and
$K(n)$ is still called the {\it sewing operation}.

We now have a family of sets $K(j)$, $j\in {\bf N}$. Given $Q\in
K(k)$, $Q_{1}\in K(j_{1}), \dots, Q_{k}\in K(j_{k})$, we define the
substitution $\gamma_{K}(Q;Q_{1}, \dots, Q_{k})$ by sucessively sewing
$Q_{i}$ to $Q$ at the $i$-th positively oriented puncture of $Q$,
$1\le i\le k$. This gives us substitution maps $\gamma_{K}$. The identity
$I_{K}$ is the conformal equivalence class of the standard sphere
${\bf C}\cup
\{ \infty\}$ with  $\infty$ the negatively oriented puncture, $0$ the only
positively oriented puncture, and with standard local coordinates
vanishing at $\infty$ and $0$. The symmetric groups $S_{j}$ acts on
$K(j)$ by permuting the orderings of the punctures of the spheres with
tubes in the conformal equivalence classes in $K(j)$ for $j\in {\bf
N}$.  The family $\{ K(j)\;|\; j\in {\bf N}\}$ equipped with these
structures forms a partial operad which we denote by $K$. Since after
a possible rescaling of the local coordinate maps the sewing operation
can always be performed, this partial operad is a ${\bf
C}^{\times}$-rescalable partial operad. It can also be shown that $K$
is associative, with associative element the equivalence class of the
sphere ${\bf C}\cup\{\infty\}$ with $\infty$ the negatively oriented
puncture, $1$ and $0$ the first and second positively oriented
punctures, respectively, and with the standard local coordinates
vanishing at these punctures. The moduli spaces $K(j)$, $j\in {\bf
N}$, have natural infinite-dimensional complex manifold structures and
the substitution maps are analytic maps. Thus $K$ is an analytic
associative ${\bf C}^{\times}$-rescalable partial operad.

The determinant line bundle over the moduli space of spheres with
boundaries induces a line bundle over $K$.  We still call it the
determinant line bundle and denote it by $\tilde{K}^{1}$. For any
complex number $c$, the line bundle $\tilde{K}^{1}$ raised to the
complex power $c$ is a well-defined line bundle over $K$ which we
denote by $\tilde{K}^{c}$. For any $c\in {\bf C}$, $\tilde{K}^{c}$ is
also an analytic associative ${\bf C}^{\times}$-rescalable partial
operad.

We consider the (partial) operad $\tilde{K}^{c}$ for a given $c\in
{\bf C}$. From the properties of the determinant line bundle, we know
that there is a natural connection on $\tilde{K}^{c}$. Moreover, this
connection is flat over $\overline{K}(j)$, the space of the conformal
equivalence classes containing the sphere ${\bf C}\cup\{\infty \}$
with the negatively oriented puncture $\infty$, the positively
oriented ordered punctures $z_{1}, \dots, z_{j-1}$, $0$, and with
standard local coordinates vanishing at these punctures, for all
$(z_{1}, \dots, z_{j-1})$, $z_{i}\ne z_{k}$ for $i<k$. In fact,
$\overline{K}(j)$ can be identified with $M^{j-1}=\{ (z_{1}, \dots,
z_{j-1})\;|\; z_{i}\ne z_{k}, \;i<k\}$.

Since an equivalence class of irreducible modules for ${\bf
C}^{\times}$ is determined by an integer $n$ such that $a\in {\bf
C}^{\times}$ acts on modules in this class as scalar multiplication by
$a^{-n}$, any completely reducible module for ${\bf C}^{\times}$ is of
the form $V=\coprod_{n\in {\bf Z}}V_{(n)}$ where $V_{(n)}$ is the sum
of the ${\bf C}^{\times}$-submodules in the class corresponding to the
integer $n$.  In particular, the vector space of a
$\tilde{K}^{c}$-associative pseudo-algebra is of this form. Note that
for a $\tilde{K}^{c}$-associative pseudo-algebra we have dim
$V_{(n)}<\infty$ by definition.

\begin{defi}
{\rm A $\tilde{K}^{c}$-associative pseudo-algebra  $(V, W,\nu)$ is  {\it
meromorphic} if the following axioms are satisfied:

(i) $V_{(n)}=0$ for $n$ sufficiently small.

(ii) For any $v'\in V'$, $v_{1}, \dots, v_{j}\in V$, $\langle v',
\nu(\cdot)(v_{1}, \dots, v_{j})\rangle$ is analytic as a function on
$\tilde{K}^{c}(j)$.

(iii) Given any $v_{1}, \dots,v_{j}\in V$ and $v'\in V'$ and any flat
section $\phi$ of the restriction of the line bundle
$\tilde{K}^{c}(j)$ to $\overline{K}(j)$, $\langle v',
\nu(\phi(\cdot))(v_{1},\dots, v_{j})\rangle$ is a meromorphic function
on $\overline{K}(j)=M^{j-1}$ with $z_{i}=0$ and $z_{i}=z_{k}$, $i<k$,
as the only possible poles, and for fixed $v_{i}, v_{k}\in V$ there is
an upper bound for the orders of the pole $z_{i}=z_{k}$ of the
functions $\langle v',
\nu(\phi(\cdot))(v_{1}, \dots, v_{i-1}, v_{i}, v_{i+1}, \dots, v_{k-1},
v_{k}, v_{k+1}, \dots, v_{j})\rangle$ for all $v_{1}, \dots, v_{i-1}$,
$v_{i+1}, \dots, v_{k-1}$, $v_{k+1}, \dots, v_{j}\in V$, $v'\in V'$ .}
\end{defi}

It can be shown that any meromorphic $\tilde{K}^{c}$-associative
pseudo-algebra is a $\tilde{K}^{c}$-associative algebra.

We call a meromorphic $\tilde{K}^{c}$-associative (pseudo-)algebra a
{\it vertex associative algebra with central charge} or {\it rank
$c$}. {\it Morphisms} (respectively, {\it isomorphisms}) of vertex
associative algebras are morphisms (respectively, isomorphisms) of the
underlying $\tilde{K}^{c}$-associative algebras.  The main theorem in
\cite{H1} and \cite{H2} can now be reformulated using the language we
have developed as follows:

\begin{theo}
The category of vertex operator algebras with central charge (or rank)
$c$ is isomorphic to the category of vertex associative algebras with
central charge (or rank) $c$.
\end{theo}

Here we give a brief description of the functor from the category of
vertex operator algebras with central charge $c$ to the category of
vertex associative algebras with central charge $c$. Let $(V,Y,{\bf
1}, \omega)$ be a vertex operator algebra with central charge $c$. The
${\bf Z}$-graded vector space $V$ is naturally a completely reducible
${\bf C}^{\times}$-module.  The module $W$ for the Virasoro algebra
generated by ${\bf 1}$ is a ${\bf Z}$-graded subspace of $V$ and
therefore is a ${\bf C}^{\times}$-submodule of $V$. In \cite{H2} and
\cite{H4}, a certain section of the line bundle $\tilde{K}^{c}$ over
$K$ is chosen. This section restricts to a nonzero flat section $\phi$
over $\overline{K}(2)$ as above.  Any element of the fiber over $Q\in
\overline{K}(2)$ is of the form $\lambda\phi(Q)$ where $\lambda \in
{\bf C}$. We define $\nu(\lambda
\phi(Q))$ by
\be
\nu(\lambda \phi(Q))(v_{1}, v_{2})
=\lambda  Y(v_{1}, x) v_{2}\lbar _{x=z}
\ee
for any $v_{1}$, $v_{2}\in V$. More generally, For $\tilde{Q}\in
\tilde{K}^{c}(2)$, $\nu(\tilde{Q})$ is obtained by modifying this
expression using exponentials of certain infinite linear combinations
of the $L(n)$ (recall (10), (11)) determined by the local coordinates
at the three punctures (see \cite{HL2}).  For $\tilde{Q}\in
\tilde{K}^{c}(j)$, $j\ne 2$, $\nu(\tilde{Q})$ can be defined analogously.
The triple $(V,W,\nu)$ is the vertex associative algebra corresponding
to $(V,Y,{\bf 1}, \omega)$.

\noindent Department of Mathematics, University of Pennsylvania,
Philadelphia,
PA 19104

\noindent email: yzhuang@math.upenn.edu

\vskip 1em

\noindent School of Mathematics, Institute for Advanced Study, Princeton,
NJ 08540

\noindent Permanent address: Department of Mathematics, Rutgers University,
New Brunswick, NJ 08903

\noindent email: lepowsky@math.rutgers.edu

\end{document}